# Substructure in Clusters of Galaxies and the value of Ω

Suvendra N. Dutta[*]
*Princeton University Observatory, Peyton Hall, Princeton, NJ 08544, USA*

**ABSTRACT**
We investigate the formation of clusters of galaxies in an expanding universe using a new code that regrids at a region of high density. In particular we investigate two models for the initial conditions, both with the standard CDM power spectrum - one has $\Omega = 1$ and the other $\Omega = 0.2$. Both universes have $H_0 = 100 \, \mathrm{km\,s^{-1}\,Mpc^{-1}}$ and $\Lambda = 0$. The level of substructure in the final cluster can be used as a discriminant of the cosmic density. We discuss various statistics which can be measured observationally from clusters of galaxies, that can be used to discriminate between the two models. We find that most of the statistics that use the clusters' member galaxies may not be the best measures of substructure. Statistics that rely more on X-ray maps and other observables depending more directly on the mass distribution could be better discriminants of $\Omega$.

## 1 INTRODUCTION

The early studies of clusters assumed them to be in equilibrium. However there has been growing evidence that clusters of galaxies have substructure. Beers & Geller (1982) used the distribution of cluster member galaxies in the sky to show that most clusters were not fully relaxed. Theije et al. (1994) have studied the shapes of clusters and found them to significantly elliptical. Large scale redshift surveys (Zabludoff et al. 1993; Bird 1994) have revealed dynamical substructure in most clusters. Using the Rosat satellite, White et al. (1993), observed significant clumpiness in the X-ray surface brightness distribution in the archetypical relaxed rich cluster, the Coma cluster. Mohr et. al. (1993) have studied the X-ray maps of several clusters and studied their axial ratio and centroid shifts, i.e., the quadrupole and dipole moments, respectively, of the intensity maps.

This substructure in clusters of galaxies is of more than morphological interest as it is potentially a probe of the mean density of the Universe. Density fluctuations continue to grow longer in an $\Omega = 1$ Universe than in a low $\Omega$ Universe. Richstone et al. (1992) pointed out that this continuance of accretion in an $\Omega = 1$ Universe means that clusters in such a Universe will have more substructure than clusters in a low $\Omega$ Universe. More recently Crone et. al. (1994) have studied the morphological dependence of clusters of galaxies on cosmological parameters. They studied the density profiles of clusters in dissipationless simulations in different cosmologies. They find the density profiles to be steeper in open universes than flat ones. Evrard et. al. (1994), studied the dissipational collapse of clusters of galaxies and found differences in the axial ratio and centroid shift of clusters in different cosmological models. Katz & White (1993) showed the presence of substructure in clusters in a simulation with dissipation.

There are a variety of tools available to quantify substructure as measured from the galaxies in the clusters. Using only redshift information Zabludoff et al. (1993) fitted the one-dimensional velocity distribution to a series expansion in Gauss-Hermite polynomials. The higher-order coefficients in this fit give the deviations of the velocity distribution from Gaussianity. Equivalent measures are the skewness and kurtosis of these one-dimensional velocity distributions. Dressler and Schectman (1988) developed a statistic that utilizes both the redshift information and distribution of the galaxies in the plane of the sky. Their statistic, $\Delta$, measures the variation of the kinematic mean and the dispersion across the cluster. Alternative statistics are the $\alpha$ statistic (West & Bothun 1990), that measures the sensitivity of the position centroids of the galaxies in the plane of the sky, to velocity range and the $\epsilon$ statistic (Bird 1994), that combines both the velocity and position information directly and uses the projected mass estimator to measure substructure.

The goal of this paper is to quantify the effectiveness of these statistics in distinguishing between open and flat Universes. Using a multi-level Particle Mesh (PM) and Tree code, we generate two sets of ten realizations and measure the substructure in the richest cluster in each of the realizations. This enables us to compute the distribution of each statistic ($\alpha$, $\Delta$, $\epsilon$, etc.) in each of the two cosmologies and to determine the degree of certainty with which an observational programme that measured 2,700 redshifts, 100 in each cluster, would distinguish between the two models. Section 2 describes the numerical method used to simulate the clusters, a hybrid PM-Tree code that has high resolution as well as large dynamic range. Section 3 discusses the various statistics used to probe substructure and introduces a statistic based on the overall mass distribution that is more appropriate to X-ray surveys. Conclusions are presented in section 4.

---

[*] current address: Dept. of Physics, Oxford University, Oxford OX1 3NP, United Kingdom.

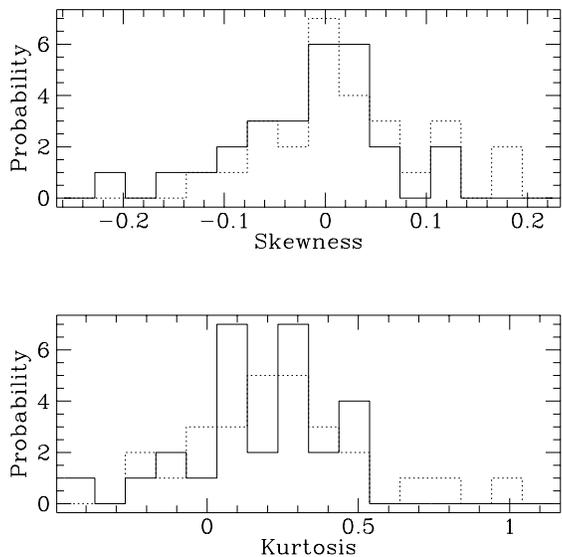

**Figure 1.** The probability distribution of the skewness and kurtosis in the radial velocity distribution. The solid line is for the Universe with $\Omega = 1$ and the dotted line for the one with $\Omega = 0.2$.

## 2 METHOD

The code used is described in detail elsewhere (Dutta 1993). Here we briefly explain the main features of the approach. We first simulate a region of the Universe, 32 Mpc on the side, using standard particle mesh (PM) code with (Park 1990; Cen 1990), with $32^3$ grid cells & particles and periodic boundary conditions. We choose a model of the Universe that has $H_0 = 100$ km s$^{-1}$ Mpc$^{-1}$ and $\Lambda = 0$. One set of simulations use $\Omega = 1$ and another set $\Omega = 0.2$. The initial density fluctuation field is a Gaussian random field with a CDM power spectrum. The power spectrum is normalised such that the $\sigma_8$ parameter (calculated in the linear approximation) is unity. We pick the highest density peak in this region and flag all PM particles that end up, or pass through, a cube 4 Mpc on the side around this peak. All these particles are broken up into 9 particles. Each has $1/9^{\text{th}}$ the mass of the original particle and they are placed on the vertices and the center of a cube of sides quarter the size of the PM grid and centered on the position of the original particle. These particles are simulated using a Tree code (Barnes & Hut 1986) kindly supplied by Dr. J. Barnes and modified to evolve the particles in the comoving coordinates. The initial conditions are realised again for these particles using a finer ($8\times$) grid. There are between $22,000 - 76,000$ tree particles in the $\Omega = 1$ models and between $17,000 - 55,000$ in the $\Omega = 0.2$ model. Now the tree and PM simulations are run again. The PM simulation doesn't know of the tree simulation. The forces on the Tree particles are twofold. One due to other Tree particles, and the other due to all the PM particles that were not broken up into Tree particles. Thus we include both mass infall and tidal forces in our simulations. We achieve the fine resolution of a particle-particle code without loss of size of the simulation box. The force softening is of the Plummer form and the softening parameter is 0.1 Mpc in comoving coordinates. It is important to make sure that the softening length of the tree particles is resolved. That is to say, we must choose our timestep to be small enough that the particles do not move more than one (r.m.s., or two to three for the particle with the highest speed) softening length per timestep. This means a timestep of 0.005 in, the scale factor, $a$. All the simulations start at a redshift of 40.

We run the simulation several times for two models. We get one realization of each Universe using the same set of random numbers and look at the same density peak in both. We have 9 such realizations for each Universe. Each realization can be treated as three because there are three independent and orthogonal viewing angles for each cluster. The two sets of clusters are compared at the final redshift (z=0).

## 3 ANALYSIS

One can divide the measures of substructure into three categories. First there are the measures that use the one-dimensional (radial) velocity distributions obtained from redshift surveys of individual clusters. If we believe Gaussian velocity distributions to be an attribute of dynamically relaxed systems one can use deviations of the velocity distribution from Gaussianity as a measure of substructure. Then there are measures that use both information on the radial velocity distribution of the galaxies and their positions in the sky. Measures such as $\Delta$ and $\alpha$ fall into this category. And finally there are the measures that do not use the galaxies but some other observable that sample the mass distribution in the cluster more smoothly. Measures of substructure found using X-ray maps or gravitational lensing would fall in this category. We should also keep in mind the kind of substructure that we are talking about. After all clusters have galaxies and that is clearly substructure. Richstone *et. al.* (1992) describe substructure as density fluctuations such that the density contrast to the local average density of the cluster isn't too large, but with enough mass in them to be more than statistical fluctuations.

### 3.1 The substructure in the radial velocity distribution

The simplest measures of substructure in the one-dimensional velocity distribution are skewness and kurtosis. Essentially the same in intent are the higher order Gauss-Hermite moments of the velocity distribution. The skewness

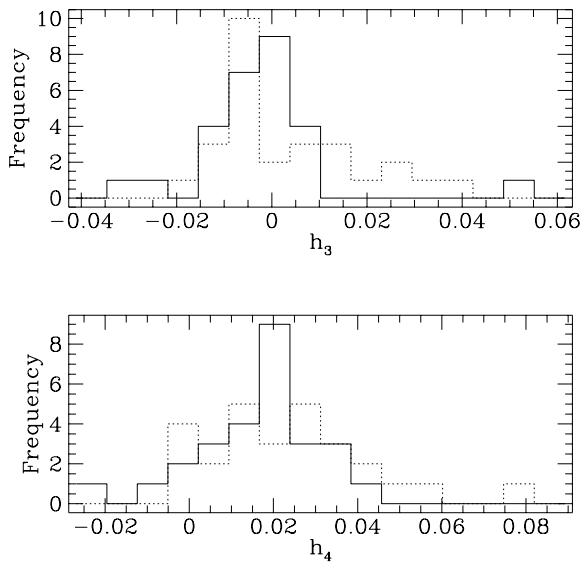

**Figure 2.** The probability distribution of the third and fourth Gauss-Hermite moments, $h_3$ and $h_4$, of the radial velocity distribution. The solid line is the model with $\Omega = 1$ and the dotted line the one with $\Omega = 0.2$.

and kurtosis of a sample of size $N_g$ are,

$$a_3 = \frac{1}{N_g \sigma_v^3} \sum_{i=1}^{N_g} (v_i - \bar{v})^3$$

$$a_4 = \frac{1}{N_g \sigma_v^4} \sum_{i=1}^{N_g} (v_i - \bar{v})^4 - 3,$$

where $\bar{v}$ is the mean and $\sigma_v$ the dispersion of the velocity distribution.

The Gauss-Hermite moments of the velocity distribution are also (Zabludoff *et al.* 1993).

$$h_l = \int_{-\infty}^{\infty} \mathcal{L}(v) \frac{e^{-w^2/2}}{\sqrt{2^{l-1} l!}} H_l(w) dv.$$

In the above $H_l(w)$ are the Hermite polynomials and $w = (v - \bar{v})/\sigma_v$ and $\mathcal{L}(v)$ is the observed velocity distribution.

Neither of these turn out to be very good measures of substructure. First of all there is no reason to think that the velocity distribution of relaxed N-body systems are Gaussian. There are many distribution functions in literature that have power-law dependancy on energy rather than an exponential. Further the nature of the substructure that ought to be expected in clusters also works against these measures. As explained before, in any hierarchical model, clusters form by the gradual accretion of subclumps that are already non-linear. Hence the velocity distribution of the cluster should not be expected to be close to Gaussian with small deviations that are functions of $\Omega$, but rather the distribution should have several components each of which may or may not be close to being Gaussian. As expected, the distributions of the skewness and kurtosis (Fig. 1) have a spread that is too wide to distinguish between the two universes. We use the K-S test to estimate the ability of the measure to distinguish between the two models. The K-S test is based on the cumulative distribution function of the samples (see Press *et. al.* 1988). The list of data points $x_i$, $i = 1, \cdots, N$ (the Gauss-Hermite moments or skewness in this case) is turned into an unbiased estimator of the cumulative distribution function of the population the sample is drawn from. This estimator, $S_N(x)$, is defined as the fraction of data points to the left of a given value x. Obviously $S_N(x)$ is a constant between consecutive $x_i$'s, and jumps by the same constant $1/N$ at each $x_i$. The Kolmogorov-Smirnov statistic $D$ is the maximum value of the absolute difference between the two estimated cumulative distribution functions. In the null hypothesis case, where the two samples are drawn from the same population the distribution of the $D$ statistic is known. We quote the probability, $P(>D)$ of getting a $D$ *higher* than the one observed given the null hypothesis. Press *et. al.* (1988), recommend accepting the null hypothesis for a sample near our size if this probabilty is less than 1%. For the skewness this probability is 52%. For the kurtosis this probability is 93%. Clearly the two models cannot be distinguished with these number of clusters or size of redshift samples per cluster. Fig. 2 shows that the same is the case with the higher Gauss-Hermite moments. In this case the difference is a little more observable. The $D$ statistic for both moments $h_3$ and $h_4$ are such that the above probability is 18%. Once again too high to distinguish between the two models. One should note that the plots show the cases where the statistics have been calculated for *all* the tree particles in the simulation. This is a sample far larger than any feasible experiment in the foreseeable future. So the deviation of the radial velocity distribution from Gaussianity is not a good discriminant of $\Omega$.

### 3.2 The substructure in the galaxy distribution

As explained before the simulations done were for dissipationless particles. This means that we do not form galaxies. We are therefore forced to making intelligent guesses for the positions of the galaxies in the cluster from the mass distribution. First the particles in the cluster are sorted by their potentials. Next the particles with the 100 largest (negetive) potential are chosen that are seperated by at least 300 kpc. These will be, roughly, the locations of the 100 deepest potential wells in cluster, *if* there are that many seperate local minima in the potential distribution. These are the logical locations of the galaxies if they could form in the simulation. For each of these potential wells the centroids of all the particles within 50 kpc of the center are calculated. The resulting mean position and velocity is taken to be the position and velocity of the representative galaxy. Although arbitrary this algorithm is acceptable because the choice of these galaxies does not appear to affect the results. We tested this by taking 100 random points in the cluster (again, seperated by at least 300 kpc) and calculated the mean position and velocity of the particles within 50 kpc for each. This produced the same results as the other algorithm.

#### 3.2.1 *The $\alpha$ statistic*

This statistic (West & Bothun 1990) measures the sensitivity of the local projected centroid of the galaxies to their radial velocities. Each galaxy is assigned a weight, $w_i = 1/\sigma_i$, which is the inverse of the velocity dispersion measured using itself and $N_{\rm kern}$ of its nearest neighbours in *velocity* space. Then the position centroids are calculated for each galaxy and its $N_{\rm kern}$ nearest neighbours in the sky using these weights,

$$x_{ci} = \frac{1}{\sum w_j} \sum_j x_j w_j$$
$$y_{ci} = \frac{1}{\sum w_j} \sum_j y_j w_j$$

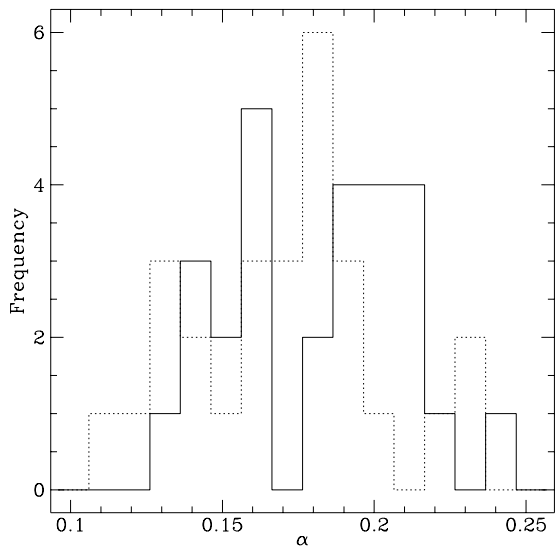
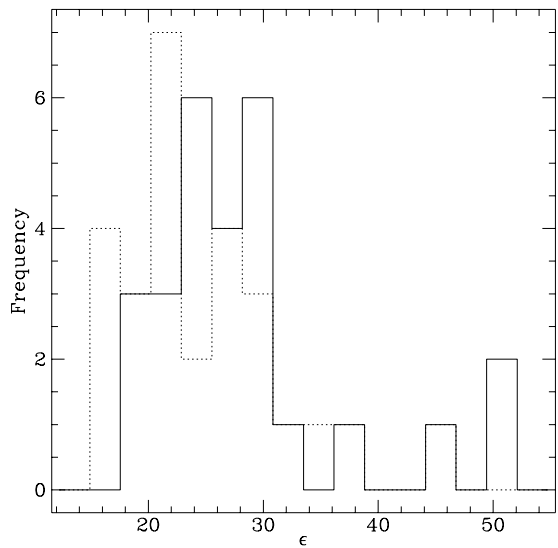

**Figure 3.** The frequency distribution of the $\alpha$ statistic. The solid line shows the case when $\Omega = 1$ and the dotted line when $\Omega = 0.2$.

**Figure 4.** The frequency distribution of the $\epsilon$ statistic. The solid line shows the case of $\Omega = 1$ and the dotted line of $\Omega = 0.2$.

From these local centroids we can define the $\alpha$ statistic as,

$$\alpha = \frac{1}{N_g} \sum_{i=1}^{N_g} \left[ (x_c - x_{ci})^2 + (y_c - y_{ci})^2 \right]^{1/2}, \qquad (1)$$

where $x_c$ and $y_c$ are the unweighted centroids for the entire sample of $N_g$ galaxies. The optimal choice for $N_{\rm kern}$ appears to be $\sqrt{N_g}$ (Bird 1994).

Fig 3. shows the distributions of the statistic in the two Universe models. As before the two world models cannot be distinguished. Using a KS test as described before we have a probability, $P(>D)$ 10%. Going through the same exercise with a 500 galaxies per cluster instead of 100 (a survey of 13,500 galaxies), we find that this probability becomes about 74%. The $\alpha$ statistic doesn't discriminate between the two models.

### 3.2.2  The $\epsilon$ statistic

This statistic mixes in information on the positions of the galaxies in the sky with their radial velocities (Bird 1994). We define a projected mass estimator, $\mathcal{M}_i$, around the $i^{\rm th}$ galaxy, using the $N_{\rm kern}$ galaxies nearest to it in the sky, as

$$\mathcal{M}_i = \left( \frac{24}{4\pi N_{\rm kern}} \right) \sum_{j=1}^{N_{\rm kern}} v_{zj}^2 r_j,$$

where $v_{zj}$ is the peculiar velocity of the $j^{\rm th}$ galaxy relative to the mean velocity measured from the $i^{\rm th}$ galaxy and ten of its nearest neighbours and $r_j$ is the distance of the $j^{\rm th}$ galaxy from the $i^{\rm th}$ galaxy. It is obvious that the $i^{\rm th}$ galaxy itself doesn't contribute to this mass. From these masses we can define the statistic

$$\epsilon = \frac{1}{M_{\rm total}} \sum_{i=1}^{N_g} \mathcal{M}_i, \qquad (2)$$

where $M_{\rm total}$ is the total mass of the cluster, calculated by adding up the masses of all the galaxies.

Fig. 4 shows the distributions of the statistic for the two different values of $\Omega$. The probability, $P(>D)$ from the K-S test is 18%. This number rises to 32% when we take 100 random points instead of our prescription. Using 500 galaxies per cluster instead of 100 the probablity gets worse to nearly 100%.

The $\epsilon$ (as well as $\alpha$ and $\delta$) statistic vanishes if the galaxies are uniformly distributed in the sky, when the estimate of the mass is the same at all points of the cluster. But naturally the galaxies aren't uniformly distributed, even for a cluster with no substructure. The cluster has a radial density profile that is falling outwards with radius. The $\epsilon$ statistic will be measuring substructure because of this non-flat radial profile. Other dissipationless simulations (Crone *et al.* 1994) indicate that the clusters have steeper radial profiles in open than in flat universes. Which could result in different distribution of $\epsilon$ measures in the two models. We can still use $\epsilon$ but now as a measure of radial profile, albeit not a very good one, rather than of substructure.

### 3.2.3  The $\Delta$ statistic

Another statistic that uses both the radial velocities as well as the positions of the galaxies in the sky is the $\Delta$ statistic (Dressler & Schectman 1988). This measure is sensitive to the fluctuations of the local kinematics at different regions of the cluster. For each galaxy we calculate the mean velocity, $\overline{v_i}$, and velocity dispersion, $\sigma_i$ from the set made up of itself and $N_{\text{kern}}$ galaxies nearest to it in the sky. These number will be different from the mean, $\overline{v_{cl}}$ and the dispersion, $\sigma_{cl}$, of the velocity distribution calculated for all the galaxies in the cluster. We can define a measure $\Delta$ that is the sum over all the galaxies of the squared difference of the local kinematics (described by $\overline{v_i}$ & $\sigma_i$) from the global kinematics (described by $\overline{v_{cl}}$ & $\sigma_{cl}$),

$$\Delta = \frac{1}{N_g} \sum_{i=1}^{N_g} \left( \frac{N_{kern}}{\sigma_{cl}^2} \right) [(\overline{v_i} - \overline{v_{cl}})^2 + (\sigma_i - \sigma_{cl})^2], \qquad (3)$$

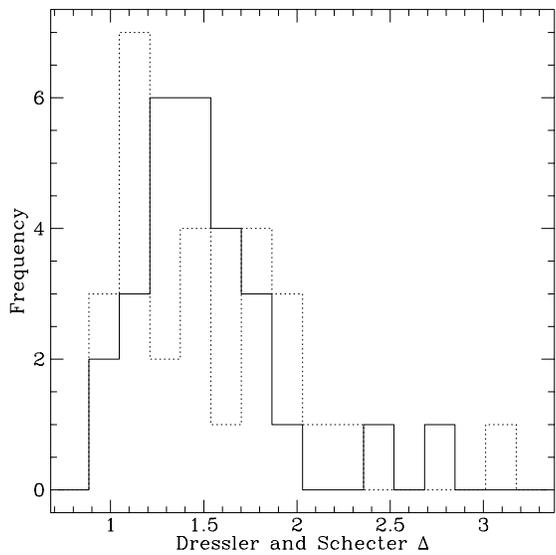

**Figure 5.** The frequency distribution of the $\Delta$ statistic. The solid line shows the model with $\Omega = 1$ and the dotted line the one with $\Omega = 0.2$.

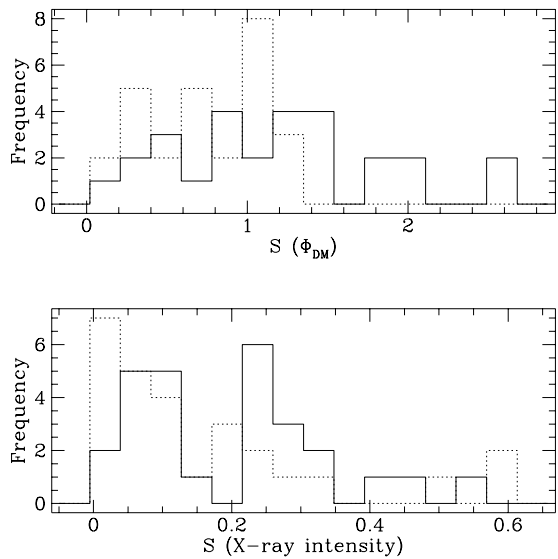

**Figure 6.** The frequency distribution of the $S$ statistic. The upper box is the statistic applied to the dark matter potential distribution and the lower box to the X-ray intensity (see discussion). The solid line shows the case of the model with $\Omega = 1$ and the dotted line the one with $\Omega = 0.2$.

Fig 5. shows the probability distribution of this statistic for the two Universe models. As can be seen there are hints of differences between the two. But the KS test shows that the probability, $P(>D)$ is 50%. With 500 galaxies instead 100 this probability goes down to 2%. So the $\Delta$ statistic is marginally discriminating between the two models, subject to the caveats of normalisation stated previously.

### 3.3 The substructure in the mass distribution

One problem in defining statistics that trace the mass distribution is that mass is not a directly observable quantity. However there are observable quantities that depend more directly on mass than galaxies. Two such observables are the X-ray maps of clusters (Mohr *et. al.* 1993; Evrard *et. al.* 1994) and gravitational lensing (Bartelmann *et al.* 1994). Since we only have dark matter particles we cannot produce X-ray maps of our clusters, but the gas in clusters should be related to the potential distribution of the dark matter matter, if dark matter dominates the gravity in clusters. There is considerable evidence (Mushotzky 1994) that the hot gas radiating X-rays in clusters are isothermal. Then the gas distribution can be approximated by the exponential of the potential distribution of the dark matter,

$$\rho_{\text{gas}} \sim e^{-\Phi_{\text{DM}}/kT_{\text{gas}}}$$

The X-ray intensity will be proportional to square of the density of gas. So we calculate the 3-D potential distribution from the particles in our simulation. From this we estimate the density of gas. We project the square of this density along one axis to get our estimate of the X-ray map of the cluster. The fact that the resulting shapes of the clusters are not unlike those in simulations using gas (Evrard *et. al.* 1993) gives the results quoted some support.

To characterize the deviation from circular symmetry we define a new statistic, the $S$ statistic. This is similar to the measures discussed in Mohr *et. al.* (1993). We use the center of the pixel with the highest X-ray intensity as the center of the map. All positions are now measured relative to this pixel. Then the $S$ statistic is defined as,

$$S = \frac{1}{I_{\text{max}}} \left| \int \mathbf{x} \, I(\mathbf{x}) \, d^2\mathbf{x} \right|, \qquad (4)$$

where, $I_{\text{max}}$ is the maximum (of the absolute value) pixel value of the potential distribution. This statistic measures the dipole moment of the intensity distribution. It was used by Mohr *et. al.*, 1993 to define substructure in cluster that indicates non-equilibrium. Evrard *et. al.*, 1993 found the same dependance of the dipole moment on $\Omega$. They find a more substructure in the flat universe because they use a lower normalisation. They also used the the ellipticity (or axial ratio) of the intensity distribution. However this need not be a measure of substructure. After all elliptical galaxies are relaxed systems. Non-vanishing even moments of the intensity distribution are expected even in relaxed systems. Non-vanishing odd moments however indicate lopsidedness and substructure. And $S$ is the first odd moment. Obviously higher order odd moments will be more sensitive measures of substructure (as well as being more prone to noise). Yet $S$ is already a good discriminant of $\Omega$. The lower box in fig. 6 shows the probability distribution of the statistic for the two Universes. The KS test yields a probability, $P(>D)$, of 10%. However it is clear from the figure that the discrimination is being spoiled by small numbers of clusters. A larger number of clusters should be more effective.

Gravitational lensing is an even more direct measure of the dark matter distribution, provided the gravitational effects of gas can be neglected. It measures substructure in the potential distribution. We can define the $S$ statistic for this using the 2-D surface distribution of potential, got by projecting the 3-D potential distribution, instead of X-ray intensity in eqn. 4. The upper box in fig. 6 shows the distribution of this statistic. The KS test on this gives a probability, $P(>D)$ less than 1%.

## 4  CONCLUSIONS

Substructure in clusters of galaxies is an inevitable prediction of hierarchical formation of structure due to gravitational collapse. The amount of substructure is a reflection of the amount of mass accumulated in the last few dynamical times. While clusters in the high $\Omega$ universe have *on average* more structure than the low $\Omega$ universe, the dispersion about the mean in each model is more bigger than the difference between the means. This makes it difficult for a observational programme to use redshift surveys of galaxies to distinguish between the two models. Our simulations suggest that surveys with samples of 100 galaxies per cluster would be insufficient to distinguish between the two models. A survey of 500 galaxies per cluster would be able to distinguish the two models on a 98% significance level. The best of these statistics appears to be the $\Delta$ statistic. However considerable caution needs to be used when using these statistics. First because they are normalised to a cluster that is uniform and real clusters have falling density profiles. And second because of the sensitiveness of the measures to Poisson noise.

The $S$ statistic is intended to probe the overall mass distribution and should be used on X-ray maps or gravitational lensing. This statistic appears to be able to distinguish between the two models with a better than 99% confidence level. The simulations described in this paper contained no gas, and hence no hydrodynamical effects were included. However other simulations (Evrard *et. al.* 1993) that did include gas appear to indicate results along the same lines as ours. But it has been seen that the X-ray maps indicate presence of a core in the gas distribution where as the gravitational lensing estimates of the mass distribution show no cores. Waxman and Miralda-Escudè (1994) show that cooling flows can lead to this. It is clear therefore that simulations with gas are needed before X-ray maps can be used with confidence. Gravitational lensing however traces only mass and dissipationless calculations like the one above should be adequate to caliberate the $S$ measure of substructure.

It ought to be noted that we have only varied $\Omega$ in the two sets of simulations. This means that the clusters in the open universe on average have considerably lower velocity dispersion than those in the flat universe. Of course all our statistics are normalised to be independant of the total mass in the cluster, however the non-linear nature of gravity will change the amount of substructure in the cluster if we change the power in the initial density fluctuation on different mass scales. This can be done in two ways. Either we can increase the amplitude of fluctuations in the open universe. Or we could increase the power on cluster scales at the cost of power on smaller scales (i.e., make the slope of the power spectrum shallower). In either case the clusters would collapse earlier and decrease further the substructure in the clusters. However the objections raised about the measures based on galaxies still stand making gravitetional lensing and X-ray maps better suited to measuring substructure.

## 5  ACKNOWLEDGMENTS

It is a great pleasure to thank Dr. David Spergel for many helpful discussions and kind guidance at all stages of this work. Many thanks are due to Dr. Binney, Dr. Hernquist and Dr. Lacey for reading through and suggesting improvements to the paper. Thanks are also due to Dr. Everard and Ms. M. Crone for many useful discussions. Helpful comments of the anonymous referee helped improve the paper. This work was funded in part by the NSF grants ASC93-18185 (HPCC GCC) and AST91-17388 as well as NASA grant NAGW-2448. Funding in Oxford was provided by PPARC.